\documentclass[12pt,english]{article}
\usepackage{times}
\usepackage[T1]{fontenc}
\usepackage[latin1]{inputenc}
\usepackage{geometry}
\usepackage{amsmath}
\usepackage{graphicx}
\usepackage{amssymb}

\makeatletter
\newcommand{\lyxaddress}[1]{
\par {\raggedright #1
\vspace{1.4em}
\noindent\par}
}

\usepackage{cite}
\usepackage{hyperref}

\usepackage{babel}
\makeatother
\begin{document}

\newcommand{\sgrad}{\boldsymbol{\nabla}_{S}}

\newcommand{\degree}{^{\circ}}

\newcommand{\dif}[1]{\vec{#1}-\vec{#1}'}

\newcommand{\dd}[2]{\delta^{#2}\left(#1-#1'\right)}

\newcommand{\vd}[2]{\delta^{#2}\left(\vec{#1}-\vec{#1'}\right)}

\newcommand{\ds}[1]{\delta\left(#1-#1'\right)}

\newcommand{\grad}{\boldsymbol{\nabla}}

\newcommand{\phik}[3]{e^{#3i#1\cdot#2}}

\newcommand{\intn}[4]{\int_{#1}^{#2}d^{n}\vec{#3}\left(#4\right)}

\newcommand{\height}{\mathcal{H}}

\newcommand{\h}{\height}

\newcommand{\f}{\mathcal{F}}

\newcommand{\dx}{\boldsymbol{\Delta}\mathbf{x}}

\newcommand{\av}{\boldsymbol{\alpha}}

\newcommand{\ve}{\boldsymbol{\eta}}

\newcommand{\ez}{\mathcal{E}_{0\degree}}

\renewcommand{\vec}{\mathbf}

\providecommand{\cs}[1][SOMETHING]{[CITE:#1]\marginpar{C} }

\providecommand{\linkable}{}

\providecommand{\bf}{\boldsymbol}

\title{Predicting and Understanding Order of Heteroepitaxial Quantum Dots.}

\author{Lawrence H. Friedman}

\maketitle

\lyxaddress{\begin{center}Pennsylvania State University, Department of Engineering
Science and Mechanics, University Park, PA 16802\par\end{center}}

\begin{abstract}
Heteroepitaxial self-assembled quantum dots (SAQDs) will allow breakthroughs
in electronics and optoelectronics. SAQDs are a result of Stranski-Krastanow
growth whereby a growing planar film becomes unstable after an initial
wetting layer is formed. Common systems are Ge$_{x}$Si$_{1-x}$/Si
and In$_{x}$Ga$_{1-x}$As/GaAs. For applications, SAQD arrays need
to be ordered. The role of crystal anisotropy, random initial conditions
and thermal fluctuations in influencing SAQD order during early stages
of SAQD formation is studied through a simple stochastic model of
surface diffusion. Surface diffusion is analyzed through a linear
and perturbatively nonlinear analysis. The role of crystal anisotropy
in enhancing SAQD order is elucidated. It is also found that SAQD
order is enhanced when the deposited film is allowed to evolve at
heights near the critical wetting surface height that marks the onset
of non-planar film growth.
\end{abstract}

\section{Introduction}

Heteroepitaxial self-assembled quantum dots (SAQDs) represent an important
step in the advancement of semiconductor fabrication at the nanoscale
that will allow breakthroughs in optoelectronics and electronics.\cite{Bimberg99,Pchelyakov2000,Grundmann2000,Petroff2001,Liu2001,Heinrichsdorff1997,Bimberg2002,Ledentsov2002,Friesen2003,Cheng2003,Krebs2003,Sakaki2003}
SAQDs are the result of a transition from 2D growth to 3D growth in
strained epitaxial films such as $\mbox{Si}_{x}$$\mbox{Ge}_{1-x}$/Si
and $\mbox{In}_{x}\mbox{Ga}_{1-x}\mbox{As}$/GaAs. This process is
known as Stranski-Krastanow growth or Volmer-Webber growth.\cite{Spencer:1991we,Bimberg99,Brunner:2002gf,Freund:2003ih}.
In applications, order of SAQDs is a key factor. There are two types
of order, spatial and size. Spatial order refers to the regularity
of SAQD dot placement, and it is necessary for nano-circuitry applications.
Size order refers to the uniformity of SAQD size which determines
the voltage and/or energy level quantization of SAQDs. It is reasonable
to expect that these type of order are linked, and it is important
to understand the factors that determine SAQD order. Further understanding
should help in the design and simulation of both spontaneous {}``bottom
up'' self-assembly and directed or guided self-assembly to enhance
SAQD order.~\cite{Shiryaev:1997iz,Kumar:fk,Friedman:2006bc,Krishna:1999ri,Hull:2003yi,Guise:patterning,Niu:2006fk,Zhao:2006qy}
Order prediction inherently involves the modeling of stochastic processes.
Recently, SAQD order has been modeled using a deterministic model
with stochastic initial conditions in the linear approximation.\cite{Friedman:fk,Friedman:2007uq}
This model was based on previous nucleationless models of SAQD formation.\cite{Spencer:1993vt,Zhang:2003tg,Liu:2003kx,Golovin:2003ms}
Here, this method of modeling SAQD order is improved by incorporating
stochastic thermal fluctuations in the surface diffusion. Thus, the
previously deterministic governing equations become stochastic. The
final order predictions are qualitatively the same as for the previous
linear model, but they are quantitatively different. Additionally,
preliminary non-linear modeling results are presented. One non-linear
model approximates a 1D surface, but incorporates the stochastic thermal
fluctuations. The second non-linear model is of a 2D surface, but
it is only implemented as a deterministic model at present. 

In the previous work using a linear deterministic model with stochastic
initial conditions\cite{Friedman:fk,Friedman:2007uq}, it was found
that peaks in the linear dispersion relation can be used to predict
and explain order. It was also found that only anisotropic models
give rise to dispersion relations with discrete peaks, thus explaining
why elastic anisotropy contributes to SAQD order as previously reported.\cite{Holy:1999th,Liu:2003kx,Liu:2003ig,Springholz:2001nx}
The dispersion relation was then used to generate a spectrum function
in the linear approximation, and the spectrum function in turn could
be used to define and predict two correlation lengths that grow as
the square root of time. These correlation lengths were identified
as the key quantities describing SAQD order. Using equations for these
correlation lengths, it was found that growth of SAQDs with an average
film height near the critical 2D-3D transition height might enhance
order, although practical limitation of producing ordered arrays of
SAQDs were also noted. Although the incorporation of a wetting potential
is possibly controversial, it appears to produce the correct phenomenology,
and it may possibly be a mathematical surrogate for more complicated
processes such as stabilization by intermixing.\cite{Tu:2004tg} See
refs.\cite{Beck:2004yq,Zhang:2003tg,Golovin:2003ms,Liu:2003kx,Spencer:1993vt,Friedman:2007uq}
for further discussion. In ref.\cite{Friedman:2007uq}, it was also
shown that the modeling/order prediction method could easily be applied
to a large class of models, but the simplest model that produced Stranski-Krastanow
growth was used as an example. Additionally, various mathematical
issues such as convergence and precise definitions of the correlation
functions as either spatial averages or ensemble averages was treated.
Readers interested in these more technical details are referred to
ref.\cite{Friedman:2007uq}.

The new result presented here is mainly the mentioned incorporation
of thermal fluctuation to seed quantum dot formation, as opposed to
the somewhat artificial assumption of a random roughness initial condition
that is chosen more or less arbitrarily. One product of the present
work is a formula to choose this initial roughness to give a nearly
equivalent disordering effect as thermal fluctuations; however, a
deterministic model will never be a true substitute for a stochastic
one. The outcomes of the stochastic model are qualitatively similar
to the previous deterministic model, but quantitatively distinct.
In addition to the stochastic linear model of SAQD order, preliminary
results of non-linear models are presented. These models appear to
corroborate the linear model predictions but also give a more complete
picture of the time evolution of SAQD order. The current model predicts
that order will be fairly poor under most growth conditions. This
seems to be in agreement with most experiments, for example refs.\cite{Baribeau:2006fj,Berbezier:2003mw,Brunner:2002gf,Gao:1999ve,Bortoleto:2003zh}.
The basic phenomenology appears to be more or less in agreement with
observations; however, more quantitative reporting of experimentally
observed order would facilitate future comparisons.

The rest of this article is organized as follows. Section~\ref{sec:Physical-Model}
presents the stochastic governing equations and physical causes of
SAQD formation. Section~\ref{sec:Linear-Stochastic-Model} presents
the linearization of the model presented in Sec.~\ref{sec:Physical-Model}
along with the extraction of order predictions and application to
growth near the critical film height using parameters appropriate
to Ge/Si SAQDs. Section~\ref{sec:Perturbatively-Non-Linear-Models}
presents preliminary non-linear modeling results. Section~\ref{sec:Conclusions}
presents the conclusion. Finally, Appendix~\ref{sec:Derivation-Ck}
presents the derivation of the time evolution equation of the spectrum
function.

\section{Physical Model\label{sec:Physical-Model}}

The formation and growth of SAQDs is modeled in a fashion similar
to refs.\cite{Spencer:1993vt,Liu:2003kx,Zhang:2003tg,Tekalign:2004jh,Friedman:2006bc}.
The film surface is described by the film height as a function of
the lateral position, $\h(\vec{x})$. The film height evolves via
surface diffusion that is driven by a diffusion potential, $\mu(\vec{x})$.
The film surface grows with a velocity normal to its surface that
is given by\begin{eqnarray}
v_{n}(\vec{x}) & = & n_{z}(\vec{x})\frac{\partial\h(\vec{x)}}{\partial t}\nonumber \\
 & = & \boldsymbol{\nabla}_{s}\cdot\left[\mathcal{D}\boldsymbol{\nabla}_{s}\mu(\vec{x})+\sqrt{2\Omega\mathcal{D}k_{b}T}\ve(\vec{x},t)\right]\dots\nonumber \\
 &  & \dots+n_{z}(\vec{x})Q,\label{eq:velocity}\end{eqnarray}
where $\mathcal{D}$ is the surface diffusivity; $n_{z}(\vec{x})$
is the $z-$component of the surface normal vector, $\hat{n}(\vec{x})$;
$\boldsymbol{\nabla}_{S}$ is the surface gradient; $\boldsymbol{\nabla}_{S}\cdot$
is the surface divergence; $Q$ is the flux of new material onto the
surface; and $\sqrt{2\Omega\mathcal{D}k_{b}T}\ve(\vec{x},t)$ is the
fluctuation of the surface diffusion. Note that the surface diffusivity
is assumed to be a scalar; thus, it is isotropic. A limited discussion
of diffusional anisotropy appears in ref.\cite{Friedman:2007uq},
and full development is in progress. The surface diffusion fluctuation
is chosen to give a steady state that is consistent with the Gibbs
distribution for a quadratic potential.\cite{Cuerno:2004wq} In ref.~\cite{Cuerno:2004wq},
there is a slope-dependent intensity factor, but here that factor
is neglected for simplicity and because it has no effect to linear
order. $\ve(\vec{x},t)$ is a time fluctuating white noise (or the
derivative of a Brownian process)\cite{Mikosch:1999vn,Gardiner:2004fk}
so that it has zero mean $\left\langle \ve(\vec{x},t)\right\rangle =0$,
where $\left\langle \dots\right\rangle $ denotes the ensemble average,
and it has a sharply peaked correlation function, $\left\langle \ve(\vec{x},t)\ve(\vec{x}',t')\right\rangle =\tilde{\mathbf{I}}\delta^{2}(\vec{x}-\vec{x}')\delta(t-t').$
$\tilde{\mathbf{I}}$ is the rank 2 identity matrix, and $\delta(x)$
is the Dirac Delta function. Eq.~\ref{eq:velocity} is interpreted
as an \^Ito stochastic differential equation.%
\footnote{Because the noise term is additive and not multiplicative, it does
not matter whether Eq.~\ref{eq:velocity} is interpreted as an \^Ito
stochastic equation or Stratonovich stochastic equation.\cite{Mikosch:1999vn,Gardiner:2004fk}%
}

The diffusion potential is derived from the total free energy. The
details of the derivation are covered in ref.\cite{Friedman:2007uq},
and only the most important points are reviewed here. The total free
energy is assumed to have two parts, elastic energy and a term that
is a combined surface energy and substrate wetting energy $\mathcal{F}=\mathcal{F}_{\text{elast.}}+\mathcal{F}_{sw}$.
The second part $\mathcal{F}_{sw}$ is an integral over the horizontal
coordinate $\vec{x}$ of an areal energy density,\[
\mathcal{F}_{sw}=\int_{\vec{x}-\text{plane}}d^{2}\vec{x}\, F_{sw}\left(\h(\vec{x}),\grad\h(\vec{x})\right).\]
The areal energy density, $F_{sw}$, is in turn a function of the
film height, $\h(\vec{x})$ and the film height gradient, $\grad\h(\vec{x})$.
From this total free energy, one can find the diffusion potential
$\mu(\vec{x})$ by taking the variational derivative with respect
to film height and multiplying by the atomic volume, $\Omega$,\begin{eqnarray}
\mu(\vec{x}) & = & \Omega\left[\omega(\vec{x})+F_{sw}^{\left(10\right)}(\vec{x})-\grad\cdot\vec{F}_{sw}^{\left(01\right)}(\vec{x})\right].\label{eq:mu2}\end{eqnarray}
$\omega(\vec{x})$ is the elastic energy density at the film surface.
$F_{sw}^{(mn)}$ indicates the $m^{\text{th}}$ derivative with respect
to $\h$ and the $n^{\text{th}}$ derivative with respect to $\grad\h$.
$F_{sw}^{\left(10\right)}(\vec{x})=\partial_{\h(\vec{x})}F_{sw}\left(\h(\vec{x}),\grad\h(\vec{x})\right)$
and each vector component of $\vec{F}_{sw}^{\left(01\right)}(\vec{x})$
is $\left[\vec{F}_{sw}^{\left(01\right)}(\vec{x})\right]_{i}=\partial_{\left[\grad\h(\vec{x})\right]_{i}}F_{sw}\left(\h(\vec{x}),\grad\h(\vec{x})\right)$.
This diffusion potential (Eq.~\ref{eq:mu2}) is a general form for
any surface diffusion model that incorporates the non-local elastic
energy density and a local areal energy density such as a surface
energy (even one with orientation dependence/faceting\cite{Zhang:2003tg})
and a wetting energy.\cite{Beck:2004yq,Zhang:2003tg,Golovin:2003ms,Liu:2003kx}
\footnote{It is possible that the wetting potential is simply an approximation
to the stabilizing effect of intermixing.~\cite{Tu:2004tg}%
} In refs.\cite{Friedman:fk,Friedman:2007uq} a simple model is analyzed
that includes elastic anisotropy, a constant surface energy density,
$\gamma$, and a substrate wetting energy density, $W(\h)$. For this
simple model, $F_{sw}=\left[1+\left(\grad\h(\vec{x})\right)^{2}\right]^{1/2}\gamma+W\left(\h(\vec{x})\right)$,
and the resulting diffusion potential is\begin{equation}
\mu(\vec{x})=\Omega\left[\omega(\vec{x})-\gamma\kappa(\vec{x})+W'(\h(\vec{x}))\right],\label{eq:musimp}\end{equation}
where $\kappa(\vec{x})$ is the total curvature, and $W'(\h)$ is
just the derivative of the wetting potential. A more extensive discussion
of different possibilities for $F_{sw}$ is discussed in ref.\cite{Friedman:2007uq}.

\section{Linear Stochastic Model\label{sec:Linear-Stochastic-Model}}

Stochastic terms that fluctuate in time lead to stochastic differential
equation that are often difficult so solve with either analytic techniques
or numerical simulation.~\cite{Gardiner:2004fk,Mikosch:1999vn} Linear
stochastic differential equations, however, are much easier to solve.
In fact, their solution is not very different from the solution of
linear deterministic (or ordinary) differential equations with stochastic
initial conditions. The linear model is naturally more approximate
than the non-linear model, but it represents an important first step,
and its solution can facilitate the development and interpretation
of non-linear models. 

To model the development of SAQD order, the growth dynamics are linearized
producing a linear dispersion relation (Sec.~\ref{sub:Linearized-Model}).
Then, the spectrum function is calculated based on the governing linear
equations and the dispersion relation (Sec.~\ref{sub:Spectrum}).
The expression for the spectrum function is then applied to the simple
diffusion potential (Eq.~\ref{eq:musimp}) for a (100) surface of
a cubic crystal (Sec.~\ref{sub:Application-to-(100)}). Application
of this method generally (to other surfaces or crystals) is outlined
in ref.\cite{Friedman:2007uq}. In this part of the calculation, crystal
anisotropy can play an important role in the diffusion dynamics and
development of SAQD order.\cite{Friedman:fk,Friedman:2007uq,Liu:2003gb,Liu:2003kx}
For simplicity, it is assumed that only elasticity has a strong anisotropic
effect. A more detailed analysis of other anisotropic effects can
be very cumbersome.\cite{Friedman:2007uq} Using the specific dispersion
relation, formulas for the correlation lengths that quantify SAQD
order and the real-space correlation function or derived. Finally,
the correlation function and the correlation lengths are applied to
a numerical example of Ge dots on a Si substrate. In this example,
order predictions and dependence of order on average film height is
compared with previous deterministic models.

\subsection{Linearized Model\label{sub:Linearized-Model}}

\begin{figure}
\begin{centering}\includegraphics[width=3.25in]{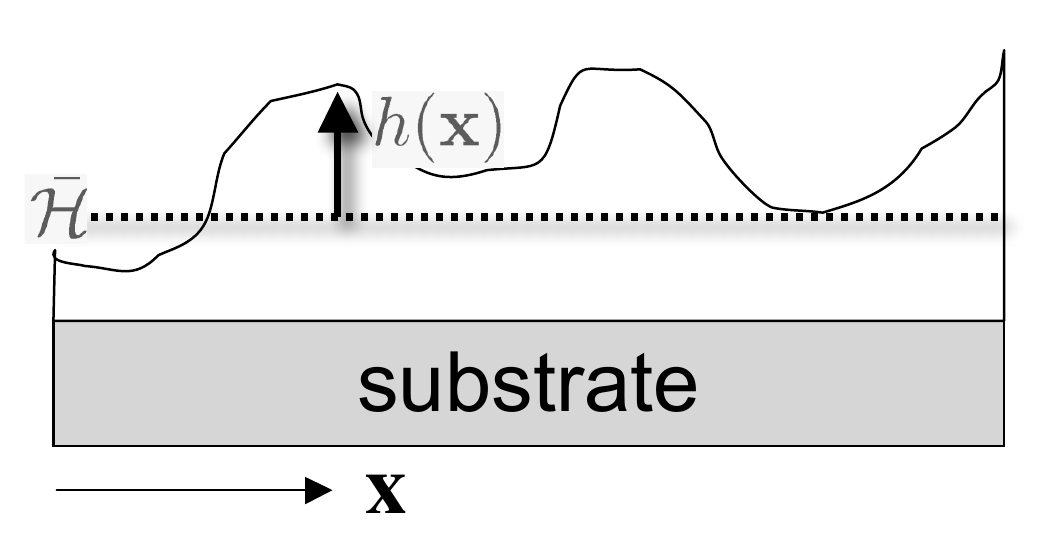}\par\end{centering}

\caption{\label{fig:Evolving-film-surface.}Evolving film surface. Total height
is average ($\bar{\h})$ plus fluctuations $h$.}
\end{figure}
Eqs.~\ref{eq:velocity} and~\ref{eq:mu2} are linearized about the
average film height (denoted $\bar{\h}$) for the case of zero deposition
rate ($Q=0$). Thus, the following analysis would correspond to a
fast deposition and then an anneal. Other growth cases such as constant
deposition rate can be analyzed in a similar fashion, but they are
beyond the scope of the present work. Following refs.\cite{Spencer:1993ve,Golovin:2003ms},
the total film height is the average film height plus small fluctuations
(Fig.~\ref{fig:Evolving-film-surface.}), \[
\h(\vec{x},t)=\bar{\h}+h(\vec{x},t).\]
Due to translational invariance of the governing equations, the Fourier
components of $h(\vec{x},t)$ evolve independently in the linear model.
Also, the non-local nature of the elastic energy makes calculations
using Fourier components (spectral methods) easier than using $h(\vec{x},t)$.
Fourier transforms use the convention, $f(\vec{x})=\int d^{2}\vec{k}\, e^{i\vec{k}\cdot\vec{x}}f_{\vec{k}}$
and $f_{\vec{k}}=(2\pi)^{-2}\int d^{2}\vec{x}\, e^{-i\vec{k}\cdot\vec{x}}f(\vec{x})$.
$h_{\vec{k}}$ is the Fourier transform of $h(\vec{x})$, where $\vec{k}$
is the corresponding wave vector.

The linearized diffusion potential is calculated following ref.\cite{Friedman:2007uq}.
Linearizing the surface-wetting part of the diffusion potential, Eq.~\ref{eq:mu2}
and taking the Fourier transform, one gets\cite{Friedman:fk}\[
\mu_{sw,\text{lin},\vec{k}}=\Omega\left(F_{sw}^{(20)}+\vec{k}\cdot\tilde{\mathbf{F}}_{sw}^{(02)}\cdot\vec{k}\right)h_{\vec{k}},\]
where the $F_{sw}^{(mn)}$ terms are the derivatives of $F_{sw}(\h,\grad\h)$,
evaluated for a perfectly flat surface of height $\bar{\h}$. They
are constants in the following analysis because they depend only on
the average film height $\h$. The first superscript indicates the
$m^{\text{th}}$ derivative of $F_{sw}$ with respect to $\h$. The
second index indicates the $n^{\text{th}}$ derivative with respect
to $\grad\h$. Evaluated for a perfectly flat surface of height $\bar{\h}$,
$F_{sw}^{\left(mn\right)}=\left.\partial_{\h}^{m}\partial_{\grad\h}^{n}F_{sw}\left(\h,\grad\h\right)\right|_{\h=\bar{\h},\,\grad\h=\vec{0}}.$
The elastic energy density at the film surface is calculated as in
refs.\cite{Friedman:2007uq,Obayashi:1998fk,Ozkan:1999gf}, where the
bimaterial (film + substrate) is approximated as an elastically homogeneous
material to simplify calculations.%
\footnote{Ref.\cite{Spencer:1993ve} treats a bimaterial, but an elastically
isotropic one.%
} The resulting elastic energy density to linear order is $\omega_{\text{lin},\vec{k}}=-\mathcal{E}_{\theta_{\vec{k}}}kh_{\vec{k}}$
so that the elastic energy is proportional to the wavenumber $k=\left\Vert \vec{k}\right\Vert $
and the Fourier component $h_{\vec{k}}$, and it has a prefactor that
depends on the wave vector direction, $\theta_{\vec{k}}$. Thus, the
total linearized diffusion potential in reciprocal space is\begin{equation}
\mu_{\text{lin},\vec{k}}=\Omega\left(-\mathcal{E}_{\theta_{\vec{k}}}k+F_{sw}^{(20)}+\vec{k}\cdot\tilde{\mathbf{F}}_{sw}^{(02)}\cdot\vec{k}\right)h_{\vec{k}}.\label{eq:mulink}\end{equation}

Linearizing the dynamic evolution, Eq.~\ref{eq:velocity}, and plugging
in $Q=0$ and $\mu_{\text{lin},\vec{k}}$,\begin{eqnarray}
\partial_{t}h_{\vec{k}}(t) & = & \sigma_{\vec{k}}h_{\vec{k}}(t)+\sqrt{2\Omega\mathcal{D}k_{b}T}\left[i\vec{k}\cdot\ve_{\vec{k}}(t)\right];\label{eq:lind1}\\
\sigma_{\vec{k}} & = & -k^{2}\mathcal{D}\Omega\left(-\mathcal{E}_{\theta_{\vec{k}}}k+F_{sw}^{(20)}+\vec{k}\cdot\tilde{\mathbf{F}}_{sw}^{(02)}\cdot\vec{k}\right),\label{eq:sd1}\end{eqnarray}
where $\ve_{\vec{k}}(t)$ is the Fourier transform of $\ve(\vec{x},t)$.
It has zero ensemble mean, $\left\langle \ve_{\vec{k}}(t)\right\rangle =0$),
and a sharply peaked two-point correlation function, $\left\langle \ve_{\vec{k}}(t)\ve_{\vec{k}'}(t')^{*}\right\rangle =(2\pi)^{-2}\delta^{2}(\vec{k}-\vec{k}')\delta(t-t').$%
\footnote{Note that because $\ve(\vec{x},t)$ is real, $\ve_{\vec{k}}(t)=\ve_{-\vec{k}}(t)^{*}$,
where {}``$^{*}$'' indicates complex conjugate. Thus,$\left\langle \ve_{\vec{k}}(t)\ve_{\vec{k}'}(t')\right\rangle =(2\pi)^{-2}\delta^{2}(\vec{k}+\vec{k}')\delta(t-t'),$
as well. %
} The growth rate of each Fourier component, $\sigma_{\vec{k}}$, is
dubbed the \emph{dispersion relation}.

\subsection{Spectrum Function\label{sub:Spectrum}}

Eqs.~\ref{eq:lind1} and~\ref{eq:sd1} can be solved as a system
of uncoupled linear stochastic ordinary differential equations with
constant coefficients\cite{Gardiner:2004fk,Mikosch:1999vn} because
the Fourier components, $h_{\vec{k}}$, evolve independently to linear
order. One could assume that there are both stochastic initial conditions
and thermal fluctuations; however, the purpose here is to analyze
the impact of just the thermal fluctuations on order. It is assumed
the film is perfectly flat at $t=0$, and that the instability is
seeded by just the thermal noise. Thus, initially, $h_{\vec{k}}(0)=0$
for all $\vec{k}$, and to linear order, the ensemble average film
height fluctuation remains zero for all time. However, the spectrum
function, $C_{\vec{k}}(t)$ provides the lowest order non-trivial
statistical description of film height fluctuations, and it is used
to predict the order of SAQD arrays in a fashion similar to refs.\cite{Friedman:fk,Friedman:2007uq}.
By taking the inverse Fourier transform of the spectrum function,
one can predict the real-space correlation function (Sec.~\ref{sub:Spectrum-and-Correlation}).
A more complete picture of the interrelations between the spectrum
function, the real-space correlation functions and other correlation
functions is presented in ref.\cite{Friedman:2007uq}.

Taking the ensemble average of Eq.~\ref{eq:lind1},\[
\partial_{t}\left\langle h_{\vec{k}}(t)\right\rangle =\sigma_{\vec{k}}\left\langle h_{\vec{k}}(t)\right\rangle +\sqrt{2\Omega\mathcal{D}k_{b}T}\left[i\vec{k}\cdot\left\langle \ve_{\vec{k}}(t)\right\rangle \right].\]
The surface diffusion thermal fluctuation is mean-zero (Sec.~\ref{sub:Linearized-Model}),
and the initial surface height fluctuation is mean-zero,$\left\langle h_{\vec{k}}(0)\right\rangle =0$;
thus, $\left\langle h_{\vec{k}}(t)\right\rangle =0$ for all time. 

Starting from the linearized governing equation and initial conditions,
an evolution equation for the spectrum function can be derived that
is both linear and deterministic (Appendix~\ref{sec:Derivation-Ck}),

\begin{equation}
\partial_{t}C_{\vec{k}}(t)=2\sigma_{\vec{k}}C_{\vec{k}}(t)+\frac{k^{2}}{(2\pi)^{2}}\left(2\Omega\mathcal{D}k_{b}T\right).\label{eq:dc}\end{equation}
Using the initial condition that $C_{\vec{k}}(0)=0$,\begin{equation}
C_{\vec{k}}(t)=\frac{\mathcal{D}\Omega k_{b}T}{(2\pi)^{2}\sigma_{\vec{k}}}k^{2}\left(e^{{2\sigma}_{\vec{k}}t}-1\right).\label{eq:sol1}\end{equation}
The spectrum function $C_{\vec{k}}(t)$ is the average value one would
expect if one extracts from a simulation or experiment the film height
power spectrum, $(2\pi)^{2}\left|h_{\vec{k}}(t)\right|^{2}/Area\approx C_{\vec{k}}(t)$.
\cite{Friedman:2007uq}

\subsection{Application to (100) surfaces\label{sub:Application-to-(100)}}

The spectrum function time dependence, Eq.~\ref{eq:sol1}, is now
applied to a (100) surfaces of cubic crystals using the simple diffusion
potential, Eq.~\ref{eq:musimp}. Anisotropy plays an important role
in order development, and for simplicity only elastic anisotropy is
included. From this analysis, the two correlation lengths are found,
and then the correlation function. Finally, a numerical example of
Ge dots on a Si substrate is presented. The dependence of order on
film height is investigated and compared and contrasted with the similar
dependence from the previous deterministic model.\cite{Friedman:2007uq}

\subsubsection{Spectrum and Correlation function\label{sub:Spectrum-and-Correlation}}

If $\sigma_{\vec{k}}$ is peaked at wave vectors, $\vec{k}_{n}$,
corresponding to some reciprocal lattice vectors, then a quasiperiodic
arrangement of SAQDs can form during the initial stages of growth.\cite{Friedman:fk,Friedman:2007uq}
This quasiperiodicity is demonstrated by applying the linearized simple
diffusion potential, Eq.~\ref{eq:musimp}, along with elastic anisotropy
$\omega(\vec{x})$ to Ge deposited on Si with a (100) substrate surface.
For a (100) surface of a crystal with cubic symmetry, $\omega_{\text{lin},\vec{k}}=-\mathcal{E}_{0\degree}\left(1-\epsilon_{A}\sin^{2}(2\theta_{\vec{k}})\right)kh_{\vec{k}}$
is a very good fit to a full elasticity calculation, where $\mathcal{E}_{0\degree}$
is the elastic energy prefactor for $\theta_{\vec{k}}=0\degree$,
and $\epsilon_{A}=\left(\mathcal{E}_{0\degree}-\mathcal{E}_{45\degree}\right)/\mathcal{E}_{0\degree}$
is an elastic anisotropy factor.\cite{Friedman:2007uq} The resulting
linear diffusion potential in reciprocal space is\cite{Friedman:2007uq}\[
\mu_{\text{lin},\vec{k}}=\Omega\left[-\mathcal{E}_{0\degree}\left(1-\epsilon_{A}\sin^{2}(2\theta_{\vec{k}})\right)k+\gamma k^{2}+W''(\bar{\h})\right]h_{\vec{k}},\]
where $\gamma$ is the surface energy density, and $W''(\h)$ is the
second derivative of the wetting potential. One can see that this
is a special case of Eq.~\ref{eq:mulink}.

The corresponding dispersion relation is\[
\sigma_{\vec{k}}=\mathcal{D}\Omega k^{2}\left[\mathcal{E}_{0\degree}\left(1-\epsilon_{A}\sin^{2}(2\theta_{\vec{k}})\right)k-\gamma k^{2}-W''(\bar{\h})\right],\]
assuming that diffusivity is isotropic as in refs.\cite{Friedman:2007uq,Friedman:fk}. 

From this dispersion relation, characteristic lengths and times can
be found along with details of the early film evolution behavior.
A characteristic wavenumber and time can be defined, $k_{c}=\mathcal{E}_{0\degree}/\gamma$
and $t_{c}=\gamma^{3}(\mathcal{D}\Omega\mathcal{E}_{0\degree}^{4})$.
Also, the strength of the wetting term $W''(\bar{\h})$ can be expressed
as a dimensionless variable, $\beta=\gamma W''(\bar{\h})/\mathcal{E}_{0\degree}^{2}$.
A detailed analysis\cite{Friedman:2007uq,Friedman:fk,Golovin:2003ms,Zhang:2003tg}
shows that a large value of $W''(\bar{\h})$ such that $\beta>1/4$
stabilizes a flat film to linear order in $h_{\vec{k}}$, while a
small value of $W''(\bar{\h})$ such that $\beta<1/4$ is insufficient
to stabilize a flat film for all possible fluctuations, $h_{\vec{k}}$.
For sufficiently small $\beta$, $\sigma_{\vec{k}}$ has 4 peaks along
the four $\left\langle 100\right\rangle $ directions, corresponding
to $\theta_{\vec{k}}=0\degree,\,90\degree,\,180\degree$ and~$270\degree$
and $k=\alpha_{0}k_{c}$. $\alpha_{0}=\left(3+\sqrt{9-32\beta}\right)/8$
is a convenient dimensionless quantity. Thus, the four peaks occur
at $\vec{k}_{1}=\alpha_{0}k_{c}\vec{i}$, $\vec{k}_{2}=\alpha_{0}k_{c}\vec{j}$,
$\vec{k}_{3}=-\alpha_{0}k_{c}\vec{i}$ and $\vec{k}_{4}=-\alpha_{0}k_{c}\vec{j}$.
Expanding $\sigma_{\vec{k}}$ in the vicinity of peak $n$,

\begin{equation}
\sigma_{\vec{k}}\approx\sigma_{0}-\frac{1}{2}\sigma_{\parallel}(k_{\parallel}-\alpha_{0}k_{c})^{2}-\frac{1}{2}\sigma_{\perp}k_{\perp}^{2},\label{eq:sigexp}\end{equation}
where $k_{\parallel}$ is the component of $\vec{k}$ parallel to
$\vec{k}_{n}$, and $k_{\perp}$ is the component of $\vec{k}$ perpendicular
to $\vec{k}_{n}$. $\sigma_{0}=\frac{1}{4}t_{c}^{-1}\alpha_{0}^{2}\left(\alpha_{0}-2\beta\right)$,
$\sigma_{\parallel}=t_{c}^{-1}k_{c}^{-2}\left(3\alpha_{0}-4\beta\right)$,
and $\sigma_{\perp}=8\epsilon_{A}\alpha_{0}t_{c}^{-1}k_{c}^{-2}$. 

Eq.~\ref{eq:sigexp} is used to find an approximate expression for
the spectrum function $C_{\vec{k}}(t)$. $\sigma_{\vec{k}}$ appears
inside an exponential; thus, for sufficiently large values $\sigma_{0}t$,
the exponential term in the vicinity of the peaks will dominate over
all other contributions to the spectrum function. Thus, $C_{\vec{k}}(t)$
will have the approximate form of four gaussians each centered around
the four peak locations, $\vec{k}_{n}$. For sufficiently narrow gaussians,
the prefactor can be approximated by its value at the peak. Thus,\begin{eqnarray}
C_{\vec{k}} & (t)\approx & \frac{\mathcal{D}\Omega k_{b}T}{(2\pi)^{2}\sigma_{0}}\left(\alpha_{0}k_{c}\right)^{2}e^{2\sigma_{0}t}\dots\nonumber \\
 &  & \dots\times\sum_{n=1}^{4}e^{-\frac{1}{2}L_{\parallel}^{2}\left(k_{\parallel}-k_{0}\right)^{2}-\frac{1}{2}L_{\perp}^{2}k_{\perp}^{2}},\label{eq:Ckanis}\end{eqnarray}
where\begin{eqnarray}
 &  & L_{\parallel}=\sqrt{2\sigma_{\parallel}t}=k_{c}^{-1}\sqrt{(6\alpha_{0}-8\beta)(t/t_{c})},\label{eq:lpar}\\
 &  & L_{\perp}=\sqrt{2\sigma_{\perp}t}=k_{c}^{-1}\sqrt{16\epsilon\alpha_{0}(t/t_{c})}.\label{eq:lperp}\end{eqnarray}

\begin{figure}
\hfill{}\includegraphics[width=3.25in]{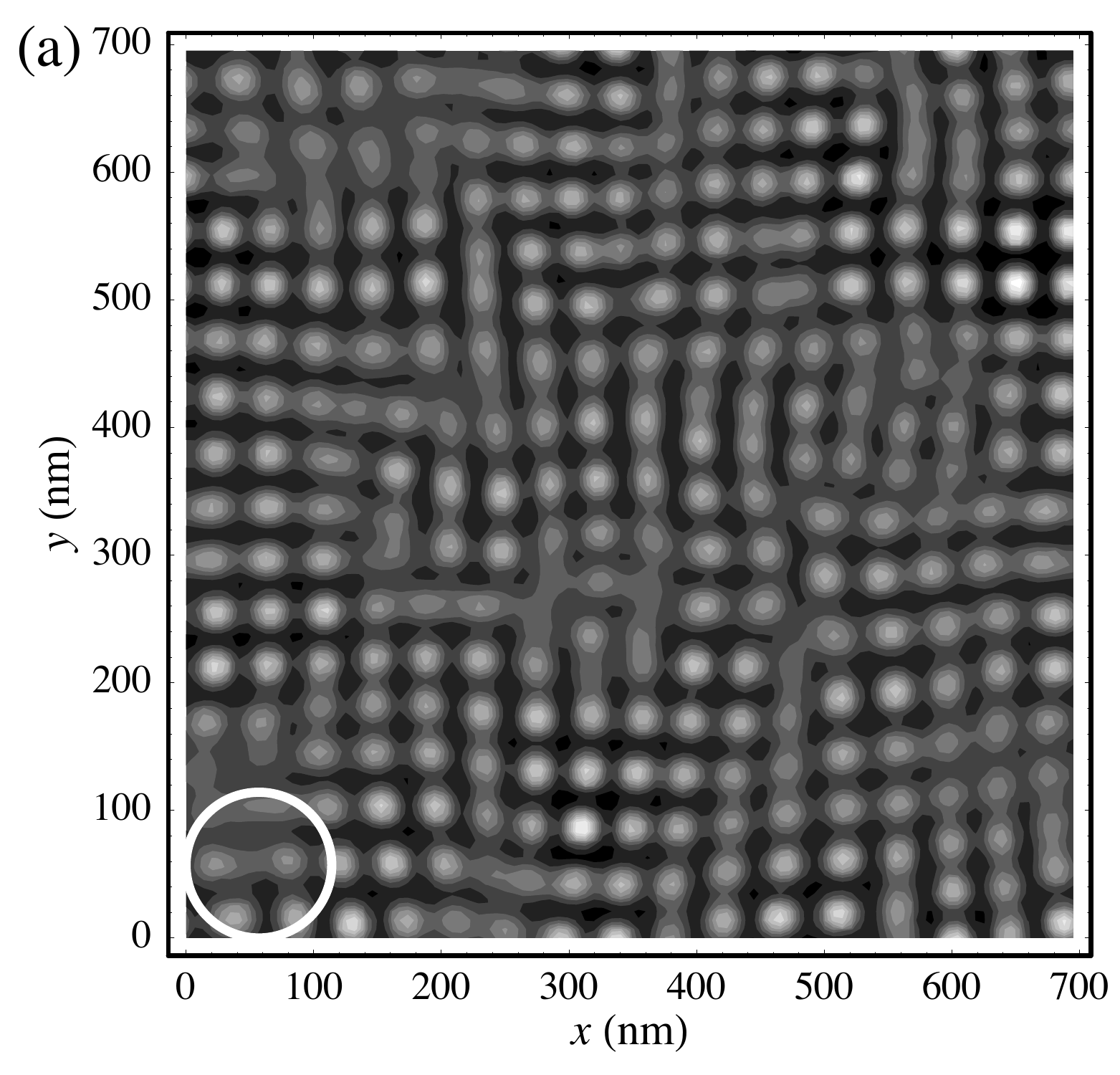}\hfill{}\includegraphics[width=3.25in,keepaspectratio]{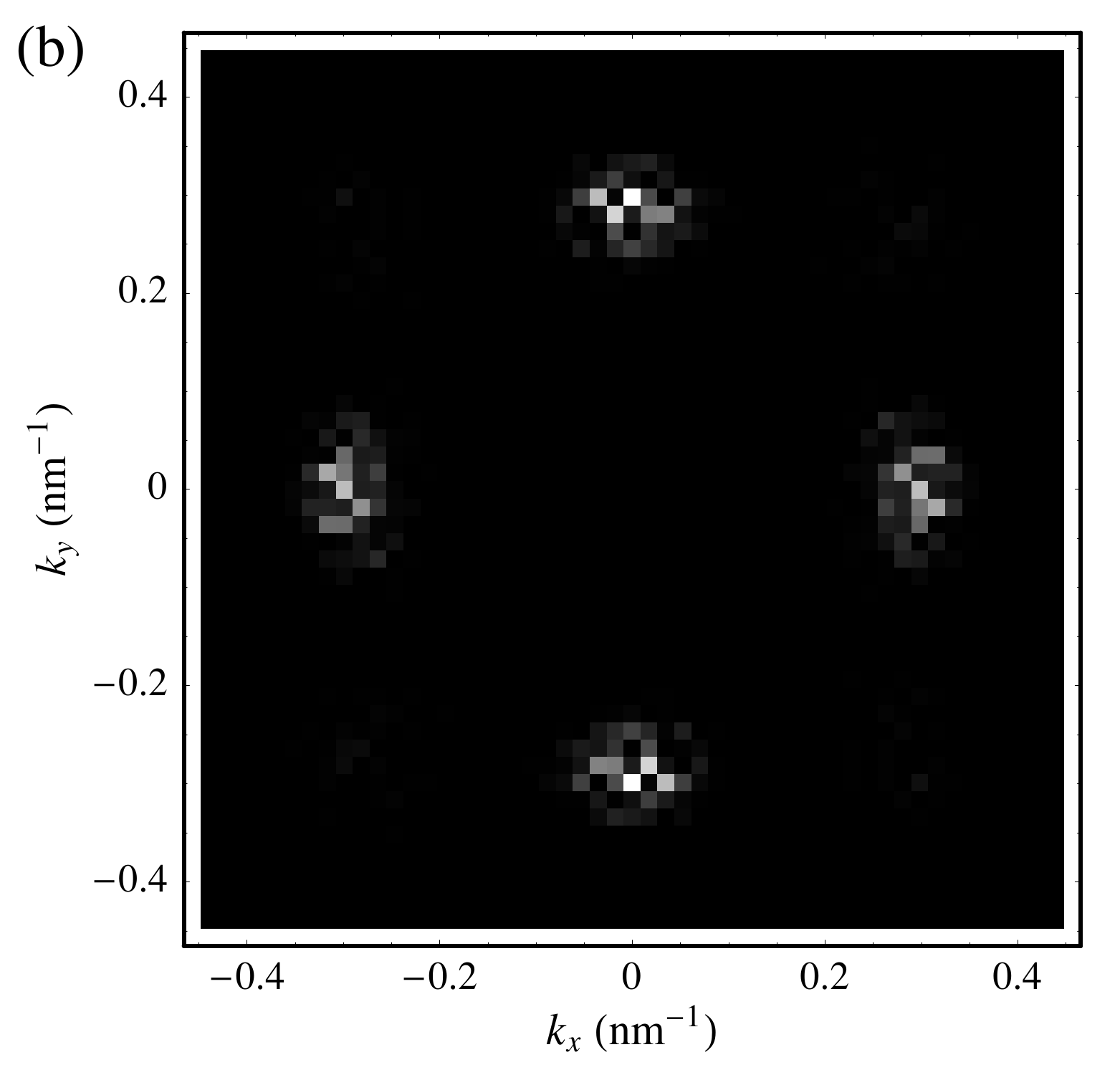}\hfill{}

\caption{\label{fig:handc}(a) Film height and (b) spectrum function of the
$\bar{\h}=1.1$$\h_{c}$ simulation discussed in Sec.~\ref{sub:2D-non-linear-deterministic}
at the end time, $t=255t_{c}$. The drawn circle in (a) has a radius
equal to $L_{\perp}$ calculated from the spectrum function (b).}
\end{figure}
$L_{\parallel}$ and $L_{\perp}$ are the two correlation lengths
that arose from models with deterministic evolution and stochastic
initial conditions. They are measures of how spatially ordered an
array of SAQDs is. The distance over which one can expect an array
of SAQDs to appear periodic is about twice the smaller of the two
correlation lengths, usually $L_{\perp}$.\cite{Friedman:fk,Friedman:2007uq}
Fig.~\ref{fig:handc} shows an example of a film surface with the
correlation length indicated. The approximate spectrum function, Eq.~\ref{eq:Ckanis},
is only valid when $\alpha_{0}k_{c}L_{\parallel}\gg1$, and $\alpha_{0}k_{c}L_{\perp}\gg1$.
Of course, when this is not the case, order will be very poor. Thus,
Eqs.~\ref{eq:Ckanis}--\ref{eq:lperp} are useful for quantifying
order when it is good, and they are able to indicate when order is
poor.

The spectrum function, Eq.~\ref{eq:Ckanis}, is very similar to the
spectrum function for the deterministic case with stochastic initial
conditions characterized by a noise amplitude $\Delta^{2}$.\cite{Friedman:fk,Friedman:2007uq}
If the noise amplitude is set to be \[
\Delta^{2}=\mathcal{D}\Omega k_{b}T\left(\alpha_{0}k_{c}\right)^{2}/\sigma_{0},\]
then the two cases are equivalent to linear order, when one performs
these similar expansions. Often, one uses deterministic evolution
equations with stochastic initial conditions as approximations to
stochastic evolution equations. By performing a suitable linear analysis
as done here, perhaps one can find an appropriate initial condition
for such approximations. Note that $\Delta^{2}$ has dimensions of
$\left[\text{length}\right]^{4}$, and the size of fluctuations in
a discretization procedure changes with the discretization length
scale. The spectral methods used here handle this problem fairly easily
as one can coarse-grain a model by simply discarding fast oscillating
noise components. A spatial discretization such as finite differencing
or the finite element method makes quantitative implementation of
white noise more complicated. 

As with the deterministic model\cite{Friedman:fk,Friedman:2007uq},
one can take the inverse Fourier transform of the spectrum function
to obtain the real-space correlation function,\begin{eqnarray}
 &  & C(\dx)=\frac{\mathcal{D}\Omega k_{b}T\left(\alpha_{0}k_{c}\right)^{2}}{\pi\sigma_{0}L_{\parallel}L_{\perp}}e^{2\sigma_{0}t}\dots\label{eq:Cxa1}\\
 &  & \dots\times\left[e^{-\frac{1}{2}\left(\Delta x^{2}/L_{\parallel}^{2}+\Delta y^{2}/L_{\perp}^{2}\right)}\cos(\alpha_{0}k_{c}\Delta x)\dots\right.\nonumber \\
 &  & \dots+\left.e^{-\frac{1}{2}\left(\Delta x^{2}/L_{\perp}^{2}+\Delta y^{2}/L_{\parallel}^{2}\right)}\cos(\alpha_{0}k_{c}\Delta y)\right]\nonumber \end{eqnarray}
The correlation function, $C(\dx)$, is a good predictor of the autocorrelation
when the sampled or simulated area is very large.\cite{Friedman:2007uq}

\subsubsection{Numerical Example and film height dependence\label{sub:Order-predictions}}

In ref.\cite{Friedman:2007uq}, it was found that for reasonably soft
wetting potentials, there can be some enhancement to spatial order
when annealing takes place for films with heights, $\bar{\h}$, that
are only just above critical film height for unstable 3D growth. This
finding was based on an assumption that the order that develops during
the initial stages of growth is a meaningful order estimate. This
assumption is justified to an extent by published numerical simulations\cite{Golovin:2003ms,Liu:2003kx,Liu:2003qi,Wang:2004dd,Ross:1998fk}
and is further justified by initial non-linear modeling results in
Sec.~\ref{sec:Perturbatively-Non-Linear-Models}. In ref.\cite{Friedman:2007uq},
the correlation lengths were found using parameters appropriate to
Ge deposited on Si. A condition for the end of the linear evolution
regime was taken to be when the r.m.s. film height fluctuation exceeded
the atomic scale, the height of one monolayer. The r.m.s. height fluctuation
is just $h_{\text{r.m.s.}}=\left[C\left(\vec{0}\right)\right]^{1/2}$,
using Eq.~\ref{eq:Cxa1}. The time at which this condition was satisfied,
$t_{\text{large}}$, was recorded, and the smaller correlation length,
$L_{\perp}$ was calculated for this time. These predicted $t_{\text{large}}$
values and the number of correlated dots in a row were graphed vs.
the dimensionless wetting potential strength for $\beta=0\dots0.25$.
It was found that the calculated time $t_{\text{large}}$ and the
calculated correlation length diverge as $\beta\rightarrow0.25$. 

\begin{figure}
\begin{centering}\includegraphics[width=3.25in,keepaspectratio]{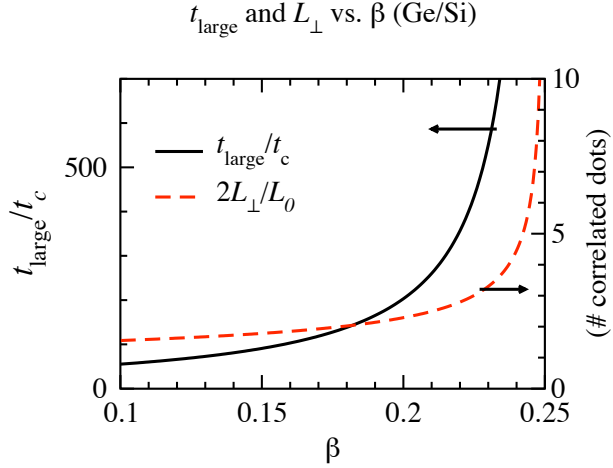}\par\end{centering}

\caption{\label{fig:tnlvb}$t_{\text{large}}/t_{c}$ and $L_{\perp}/L_{0}$
vs. the dimensionless wetting parameter $\beta$ for Ge/Si at $T=600\degree\text{C}$
as discussed in Sec.~\ref{sub:Order-predictions}}
\end{figure}
The same procedure is now followed for the present model for Ge on
Si with temperature $T=600\degree\text{ C}$. All values are the same
as for the calculations in ref.\cite{Friedman:2007uq}. The results
are graphed in Fig.~\ref{fig:tnlvb}. When compared with the results
from the deterministic model with stochastic initial conditions\cite{Friedman:2007uq},
one finds that the effect of thermal fluctuations in the surface diffusion
are actually more disruptive to order than assuming an initial surface
with atomic scale roughness. The qualitative trends are the same,however,
and the divergence in correlation length as $\beta\rightarrow0.25$
is observed. As discussed in ref.\cite{Friedman:2007uq}, one should
take care interpreting this result, and there are of course practical
limitations. The order enhancing effect of near critical growth has
not been experimentally observed (or looked for), and there may be
practical limitation to implementing near critical growth as a method
to enhance order such as the requirement for precise deposition control.

\section{Perturbatively Non-Linear Models\label{sec:Perturbatively-Non-Linear-Models}}

The order estimates presented in refs.\cite{Friedman:fk,Friedman:2007uq}
and Sec.~\ref{sub:Order-predictions} are based on the order that
develops before fluctuations become large. The significance of these
calculations is based on the following observations:

\begin{enumerate}
\item Order increases during the linear stage of growth as $t^{1/2}$ (Eqs.~\ref{eq:lpar}
and~\ref{eq:lperp}).
\item Order does not increase forever. If it did, growing perfectly ordered
arrays of dots would be trivial. Also, qualitative analysis of numerical
simulations bears this out.~\cite{Golovin:2003ms,Liu:2003kx,Liu:2003qi,Wang:2004dd,Ross:1998fk}
\end{enumerate}
It is, of course, worthwhile to extend the method of quantifying and
predicting order to non-linear models. Non-linear stochastic modeling
can be very cumbersome and difficult to implement, but some preliminary
results are presented here. The same system as in Sec.~\ref{sub:Order-predictions}
and ref.\cite{Friedman:2007uq} is modeled here, and the same parameters
are used.

\subsection{1D Multiscale-Multitime Expansion\label{sub:1D-Multiscale-Expansion}}

First, the results of a 1D non-linear model with stochastic evolution
is presented. As a first attempt at non-linear modeling, two approximations
are made. First, the elastic and surface energy parts are completely
linearized. Second, the wetting potential, $W(\h)$ is treated using
a multiscale-multitime expansion.~\cite{Golovin:2003ms,Cross:1993ti}
Full details of the model are omitted out of space considerations
and because these are preliminary results.

\begin{figure}
\begin{centering}\includegraphics[width=3.25in]{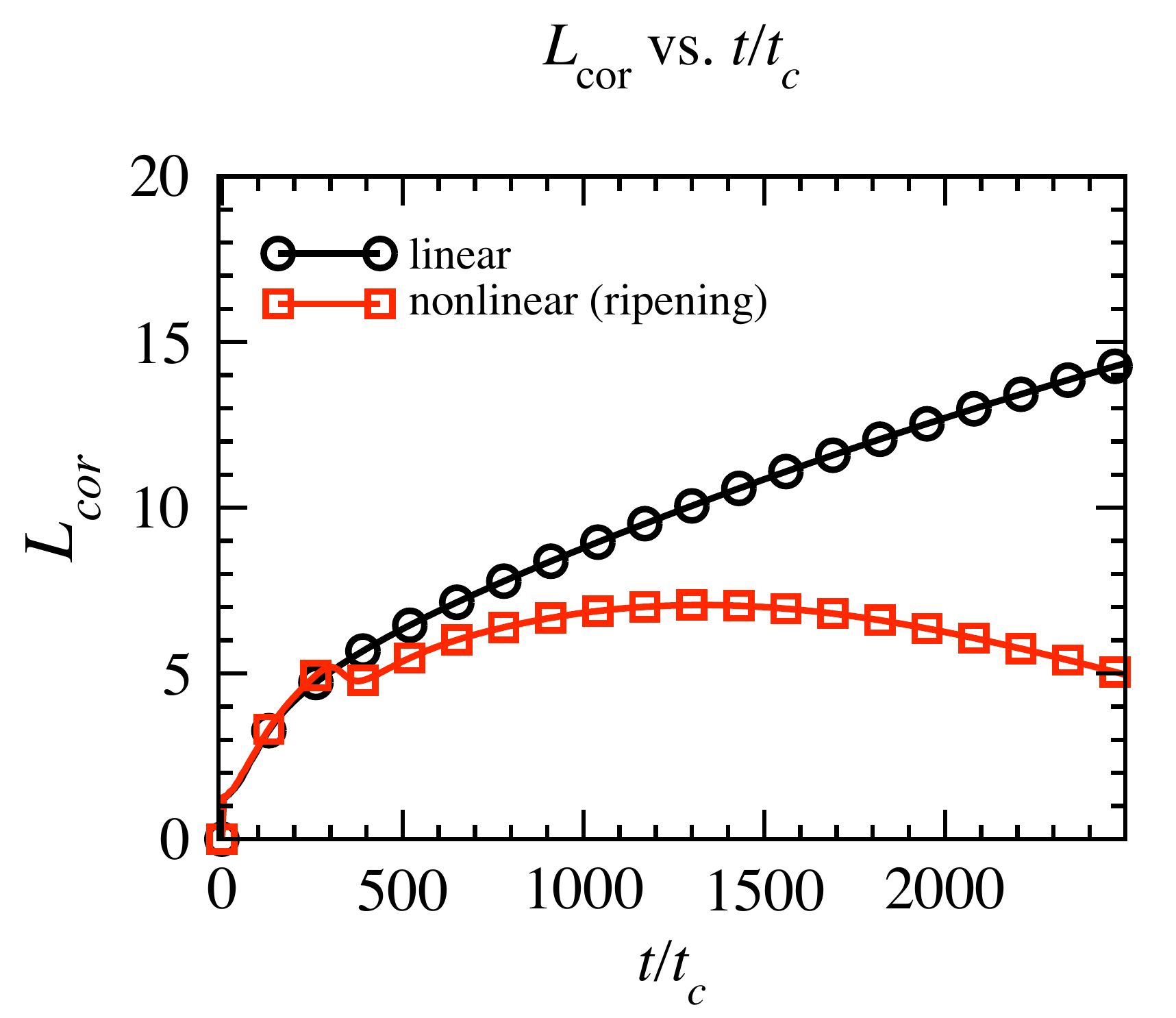}\par\end{centering}

\caption{\label{fig:1Dres}Time dependence of number of correlated dots $L_{\text{cor}}/L_{0}$
vs. dimensionless time, $t/t_{c}$ for 1D stochastic film evolution
(Sec.~\ref{sub:1D-Multiscale-Expansion}). Both linear model and
multiscale expansion results are shown.}
\end{figure}
Based on ref.\cite{Zhang:2003tg}, the wetting potential is chosen
to be $W(\h)=2.314\times10^{-6}/\h\text{ erg/cm}^{2}$ with $\h$in
cm. This gives a critical film height of 4 monolayers $=1.132\text{ nm}$.
The simulated film has an average film height of $\bar{\h}=1.358\text{ nm}$
giving $\beta=0.1447$. The simulation cell size is $19.68\,\mu\text{m}$,
large enough to hold 513 dots of average size $L_{0}=2\pi/k_{0}=38.4\text{ nm}$.
The multiscale-multitime expansion uses an expansion in a scale variable
$\epsilon$ to create a perturbation-like series. Additionally, fast
oscillating Fourier components of $W(\h)$ are discarded so that the
natural length scale is the average size of a single dot, $L_{0}$
. To fourth order in $\epsilon$, one obtains a set of two coupled
partial differential equations.\cite{Golovin:2003ms,Cross:1993ti}
These equations are solved using Stochastic Euler numerical integration\cite{Mikosch:1999vn,Gardiner:2004fk}
implemented with Mathematica.\cite{mathematica_5.2} Computational
efficiency is greatly enhanced by the multiscale-multitime expansion,
but of course, accuracy and fidelity to the original model is partially
sacrificed. Correlation lengths are calculated from the peak widths
of the spectrum function ($\Delta k$), using $L_{\text{cor}}=1/\Delta k$.
The number of dots in a row that form a recognizably periodic structure
is $\# dots=2L_{\text{cor}}/L_{0}$. The time evolution of this number
is plotted for both the linear model and the stochastic simulation
(Fig.~\ref{fig:1Dres}). The linear model has a correlation length
that grows indefinitely. The non-linear model has a correlation length
that grows to a peak value and then shrinks. In this case, the onset
of ripening ruins the SAQD order. The onset of ripening in this model
corresponds to the {}``blow-up solution'' in ref.\cite{Golovin:2003ms}

\subsection{2D non-linear deterministic model\label{sub:2D-non-linear-deterministic}}

\begin{figure}
\begin{centering}\includegraphics[width=3.25in]{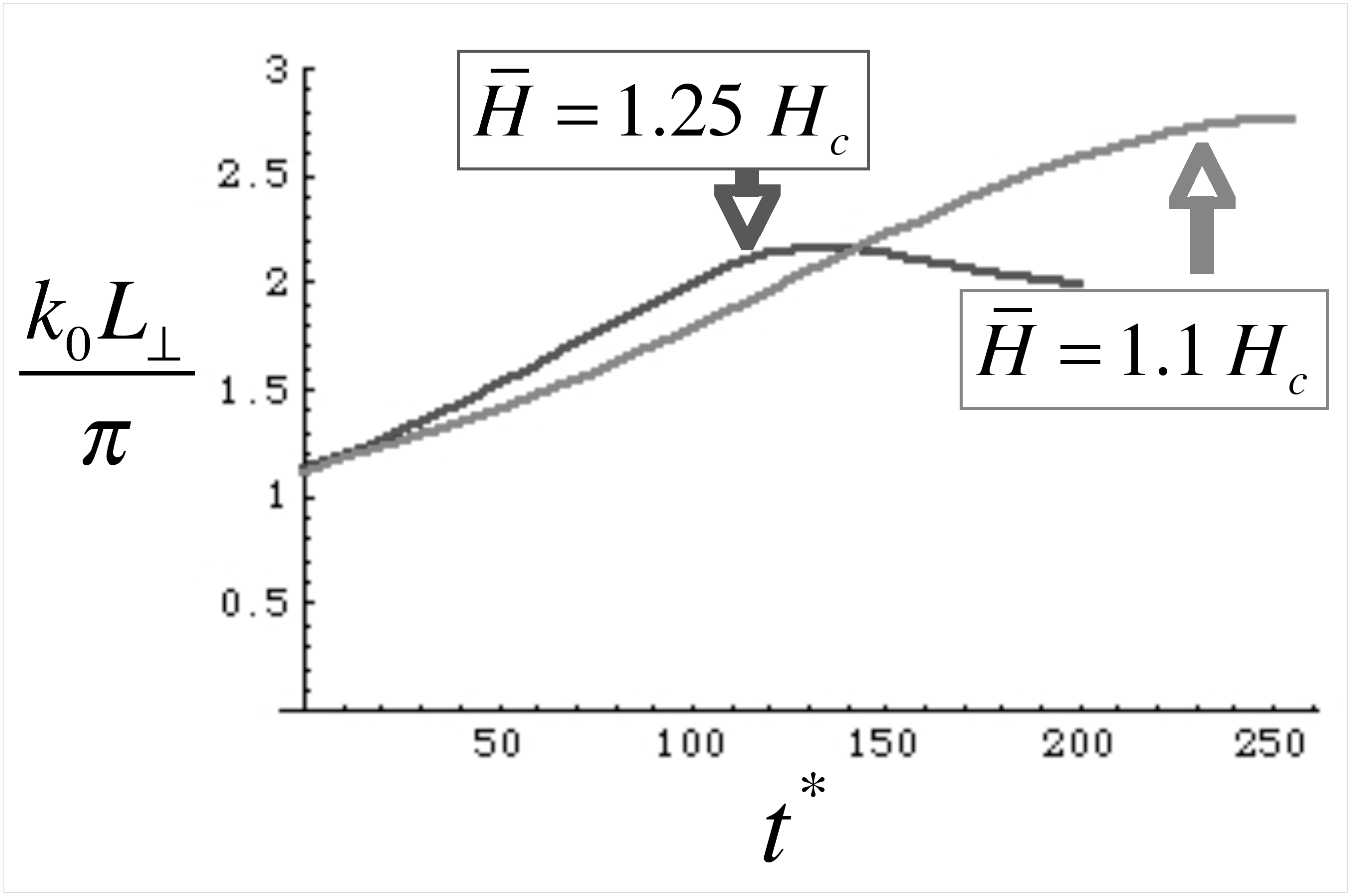}\par\end{centering}

\caption{\label{fig:2D_cor_L}Dimensionless correlation length vs. dimensionless
time for 2D non-linear deterministic model with stochastic initial
conditions. Results for two different average film height are reported,
$\bar{\h}=1.25\h_{c}$ and $\bar{\h}=1.1\h_{c}$}
\end{figure}
A similar result is obtained for a 2D deterministic non-linear model.
This model treats the surface energy and wetting energy in full non-linear
fashion. The non-local elastic part is found to cubic order in the
film height fluctuation in $h$ via a perturbation series. The stochastic
initial conditions are sampled white noise with an initial atomic
scale roughness, corresponding to $\Delta^{2}=0.0403\text{ nm}^{4}$.\cite{Friedman:2007uq}
The critical height for the 2D-to-3D-growth transition is $\h_{c}=1.132\text{ nm}$.
Two initial average film heights are used to investigate the trend
predicted in Fig.~\ref{fig:tnlvb}, $\bar{\h}=1.25\h_{c}=1.415\text{ nm}$
($\beta=0.1280$) and $\bar{\h}=1.1\h_{c}=1.245\text{ nm}$ ($\beta=0.1878$).
The simulation cell size corresponds to 17 dots squared = 289 dots.
The time evolution equations are solved using the native numerical
differential equation solver in Mathematica.\cite{mathematica_5.2}
The correlation lengths vs. dimensionless time are plotted for both
cases in Fig.~\ref{fig:2D_cor_L}. In both cases, the correlation
length increases early on while fluctuations are small, reaches a
peak value and then decreases due to ripening. The peak value of the
correlation length is greater for the second case with $\beta$ closer
to the optimal value of $1/4$. The 2D non-linear deterministic model
further substantiates the theory that order develops during the early
growth stages and then is diminished during ripening. Furthermore,
the trend predicted by the linear order model is it least qualitatively
corroborated because growth near the critical threshold enhances the
peak order of SAQDs according to the 2D non-linear model.

\section{\label{sec:Conclusions}Conclusions}

A linear stochastic model of SAQD order has been presented as an extension
of a previous linear deterministic model of the order of epitaxial
self-assembled quantum dots (SAQDs). In addition, some preliminary
results from non-linear stochastic and non-linear deterministic models
have been presented to substantiate the significance of, extend and
clarify the linear models of SAQD order. The presented numerical examples
were based on a very simple SAQD model, and there has been much advancement
in SAQD growth modeling; however, the presented procedure should apply
equally well to a wide variety of models with various phenomenological
assumptions and help to augment them and quantitatively extract order
predictions. The current stochastic model should be augmented in the
future to reflect these advances. Some adaptation of the method ought
to apply to attempts to engineer SAQD order as well, such as substrate
patterning or growing multilayers of SAQDs. As with the previous deterministic
model, two correlation lengths are found, longitudinal $L_{\parallel}$
and transverse $L_{\perp}$ . The transverse correlation length appears
to be the most limiting, and thus should be used to estimate order.
It is found that if a wetting potential is incorporated that is sufficiently
soft, growth near the 2D-3D transition critical film height enhances
SAQD order; however, this enhancement would require very precise experimental
control to implement. Nevertheless, it demonstrates how the presented
methods might apply to other attempts to optimize SAQD growth and
could help engineer those processes. It was also found that the previous
deterministic model can be made approximately equivalent to the present
stochastic model by choosing the appropriate initial conditions.  Preliminary
non-linear modeling appears to corroborate these claims, at least
qualitatively. A quantitative comparison is still needed. The method
to extract SAQD order should help with phenomenological model development
as the correlation lengths and possibly other statistical characterization
should facilitate quantitative tuning of phenomenological models to
experiments. The models presented here apply to the nucleationless
mode of SAQD formation; however, the inclusion of thermal fluctuations
in non-linear models should facilitate a conceptual and/or mathematical
unification of models of SAQD thermal nucleation and the nucleationless
mode.

Thanks to L. Fang for a critical reading of this manuscript.

\appendix

\section{\label{sec:Derivation-Ck}Derivation of Eq.~\ref{eq:dc}}

The two-point correlation function in reciprocal space is\[
C_{\vec{k}\vec{k}'}(t)=\left\langle h_{\vec{k}}(t)h_{\vec{k}'}(t)^{*}\right\rangle .\]
Note that at time $t=0$, $C_{\vec{k}\vec{k}'}(0)=\left\langle h_{\vec{k}}(0)h_{\vec{k}'}(0)^{*}\right\rangle =0$.
The time evolution of $C_{\vec{k}\vec{k}}(t)$ can be found using
the stochastic chain rule ($\hat{\text{I}}$to's lemma) and then taking
the ensemble average.\cite{Mikosch:1999vn,Gardiner:2004fk}\begin{eqnarray}
\partial_{t}C_{\vec{k}\vec{k}'}(t) & = & \left(\sigma_{\vec{k}}+\sigma_{\vec{k}'}\right)C_{\vec{k}\vec{k}'}(t)\dots\label{eq:dckk}\\
 &  & \dots+\frac{\vec{k}\cdot\vec{k}'}{(2\pi)^{2}}\left(2\Omega\mathcal{D}k_{b}T\right)\delta^{2}(\vec{k}-\vec{k}').\end{eqnarray}
The thermal fluctuations only contribute if $\vec{k}=\vec{k}'$. Since
initially $C_{\vec{k}\vec{k}'}(0)=0$, one can expect $C_{\vec{k}\vec{k}'}(t)$
to be non-zero only if $\vec{k}=\vec{k}'$. Thus, the two-point correlation
function is determined completely by the ensemble averaged spectrum
function, $C_{\vec{k}}(t)$ as in ref.\cite{Friedman:2007uq},\begin{equation}
C_{\vec{k}\vec{k}'}(t)=C_{\vec{k}}(t)\delta^{2}(\vec{k}-\vec{k}').\label{eq:rp}\end{equation}
This results is only strictly true for the linearized equation. From
Eq.~\ref{eq:dckk} the time evolution equation of the spectrum function
is found by inspection to be Eq.~\ref{eq:dc}.

\bibliographystyle{unsrt}
\bibliography{jem_tms_abbr}

\end{document}